\documentclass[
 amsmath,amssymb,
 aps,
prb,
showpacs
]{revtex4}

\usepackage{graphicx}
\usepackage{dcolumn}
\usepackage{bm}

\begin{document}


\title{Novel properties of graphene in the presence of energy gap: \\ 
optics, transport and mobility studies} 

\author{Godfrey Gumbs$^{1,2}$}
\affiliation{$^{1}$ Department of Physics and Astronomy, Hunter College at the \\
City University of New York, \\
695 Park Avenue, New York, NY 10065, USA \\
 \\
$^2$Donostia International Physics Center (DIPC), P de Manuel Lardizabal, 4, \\
20018, San Sebastian, Basque Country, Spain}

\author{Danhong Huang}
\affiliation{Air Force Research Laboratory, Space Vehicles Directorate, \\
Kirtland Air Force Base, NM 87117, USA
}
 
\author{Andrii Iurov\footnote{\texttt{E-mail contact}: theorist.physics@gmail.com} and Bo Gao}
\affiliation{
  Department of Physics and Astronomy, Hunter College at the \\
City University of New York, \\
695 Park Avenue, New York, NY 10065, USA 
}
\date{\today}

\pacs{73.40.Gk, 61.48.-c, 72.10.Fk, 72.20.Ht, 05.45.-a, 73.43.Lp, 78.30.Na}

\begin{abstract}
We review the transmission of Dirac electrons through a potential barrier in the presence of circularly polarized light. A different type of transmission is demonstrated and explained. Perfect transmission for nearly head-on collision in infinite graphene is suppressed in gapped dressed states of electrons. We also present our results on enhanced mobility of hot Dirac electrons in nanoribbons and magnetoplasmons in graphene in the presence of the energy gap. The calculated carrier mobility for a graphene nanoribbon as a function of the bias field possesses a high threshold for entering  the nonlinear transport regime. This threshold is a function of  both extrinsic and intrinsic properties, such as lattice temperature, linear density, impurity scattering strength, ribbon width, and correlation length for the line-edge roughness. Analysis of  non-equilibrium carrier distribution function confirms  that the difference between linear and nonlinear  transport is due to sweeping electrons from the right  to left Fermi one through  elastic scattering as well as  moving electrons from low to high-energy ones through field-induced heating. The plasmons, as well as the electron-hole continuum are determined by both energy gap and the magnetic field, showing very specific features, which have been studied and discussed in details. 
\end{abstract}
\maketitle


\section{Introduction}

A considerable amount of interest in basic research and device
development has been generated for both the electronic
and optical properties of two-dimensional (2D) graphene material\,\cite{dh-ref-1.4,dh-ref-2.2,dh-ref-2.3,dh-ref-2.4,dh-ref-2.5}.
This began with the first successful isolation of single graphene layers and
the related transport and Raman experiments for such layers\,\cite{dh-ref-2.6}.
It is found that the major difference between a graphene sheet and a
conventional 2D electron gas (EG) in a quantum well (QW) is
the band structure, where the energy dispersions of
electrons and holes in the former are linear in momentum space,
but quadratic for the latter. Consequently, particles in graphene
behave like massless Dirac fermions and display many unexpected
phenomena in electron transport and optical response, including the anomalous
quantum Hall effect\,\cite{dh-ref-1.2,dh-ref-2.8}, bare and dressed
state Klein tunneling\,\cite{dh-ref-2.9,dh-ref-2.10,dh-ref-2.11,dh-ref-2.12}
and plasmon excitation\,\cite{dh-ref-2.13,dh-ref-2.14,dh-ref-2.15},
a universal absorption constant\,\cite{dh-ref-2.16,dh-ref-2.17}, tunable intraband\,\cite{dh-ref-2.18} and
interband\,\cite{dh-ref-2.19,dh-ref-2.20} optical transitions, broadband $p$-polarization effect\,\cite{dh-ref-2.21}, photo-excited hot-carrier thermalization\,\cite{dh-ref-2.22}
and transport\,\cite{dh-ref-2.23}, electrically and magnetically tunable band structure for ballistic transport\,\cite{dh-ref-2.24}, field-enhanced mobility in
a doped graphene nanoribbon\,\cite{dh-ref-1.0} and electron-energy
loss in gapped graphene layers\,\cite{dh-ref-2.26}.

Most of the unusual electronic properties of graphene may be
explained by single-particle excitation of electrons. The Kubo
linear-response theory\,\cite{dh-ref-2.27} and Hartree-Fock theory
combined with the self-consistent Born
approximation\,\cite{dh-ref-2.28} were applied to diffusion-limited
electron transport in doped graphene. Additionally, the semiclassical
Boltzmann theory was employed for studying transport in both linear\,\cite{dh-ref-1.16} and nonlinear\,\cite{dh-ref-1.0} regimes.
For the plasmon excitation in graphene, its important role in
the dynamical screening of the electron-electron interaction\,\cite{dh-ref-2.13,dh-ref-2.14,dh-ref-2.15,dh-ref-2.26,dh-ref-2.30}
has been reported. However, relatively less attention has been received for the electromagnetic (EM) response of graphene materials, especially
for low-energy intraband
optical transions\,\cite{dh-ref-2.18,dh-ref-2.31,dh-ref-2.32}.

Gapped graphene has marked an important milestone in the study of
graphene's electronic and transport properties from both a theoretical
and experimental point of view as well as in  practical device
applications. The reason for this is that
gapped graphene has  applications  such as a field-effect
transistor where a gap is essential as well as graphene interconnects.
The effective band gap may be generated by  spin-orbit interaction,
or when monolayer graphene is placed on a substrate such as
ceramic silicon carbide  or graphite. The gap may also arise
dynamically when  graphene is irradiated  with circularly polarized
light. Depending on the nature of the substrate on which graphene
is placed or the intensity or amplitude of the light, the gap may be a
few meV or as large as one eV\,\cite{bib:AI:Kibis}. In general, the energy gap
is attributed to a breakdown in symmetry between the sublattices
caused by external perturbing fields from the substrate or photons
coupled to the atoms in the $A$ and $B$ sublattices.

This Section will be divided into two parts covering the electrical and optical
properties of electrons in gapped graphene materials. The first part
will deal with  electrical transport. Topics will include: (1) unimpeded
tunneling of chiral electrons in graphene nanoribbons; (2) anomalous
photon-assisted tunneling of Dirac electrons in graphene; and (3) field
enhanced mobility by nonlinear phonon scattering of Dirac electrons.
The second part will focus on the collective plasmon (charge density)
excitations  for:   intraband and interband plasmons.

\section{Dirac Fermions in Graphene: Chirality and Tunable Gap}

Graphene, an allotrope of carbon, which has been recently discovered in experiment\,\cite{dh-ref-1.2}, has become one of the most important and
extensively studied materials in  modern condensed matter physics,
mainly because of its extraordinary electronic and transport properties\,\cite{bib:AI:Katsnelson}. According to Ref.\,\cite{bib:AI:geim:2009}, graphene can be described as a single atomic plane of graphite, which
is sufficiently isolated from its environment and considered to be
free-standing. The typical carbon-carbon distance in a graphene layer is $0.142$\,nm, and the interlayer distance in a graphene stack is $0.335$\,nm.
Any graphene sample with less than $2.4 \times10^4$ carbon atoms or
less than $20$\,nm of length is unstable\,\cite{bib:AI:structure},
tending to convert to other fullerenes or carbon structures.
Now, we will briefly discuss the most crucial electronic properties
of graphene, relevant to the electron tunneling phenomena.
A good description of the principal electronic structure and
properties may be found in Refs.\,\cite{dh-ref-1.4,bib:AI:DSTranspG}.
Surprisingly, the first theoretical study of ``graphene'' was performed
more than sixty years ago\,\cite{bib:AI:Wallace1947,bib:AI:mcclure}.
The most striking difference in comparison with conventional semiconductors or metals is the fact that low-energy electronic
excitations in graphene are massless with Dirac cones for energies.
The quite complicated energy band structure of graphene may be approximated
as a cone (\textit{Dirac cone}) in the vicinity of the two inequivalent
corners, i.e., $K$ and $K^{\prime}$ points, of the Brillouin zone.
In summary, the electronic properties of graphene may be approximately
described by the \textit{Dirac equation}, corresponding to the linear energy dispersion  next to the $K$ and $K^{\prime}$ points:

\begin{equation}
-i\hbar v_F\,\sigma\cdot\mbox{\boldmath$\nabla$}
\Psi({\bf r})=\varepsilon\,\Psi({\bf r})\ ,
\label{AI:eq:1}
\end{equation}
where $v_F=c/300$ is the Fermi velocity.  This form is similar to
the high-energy quantum electrodynamics (QED) Dirac equation.
In  momentum (${\bf k}$) space, the Hamiltonian is simplified as follows:

\begin{equation}
\hat{\mathcal{H}}=\hbar v_F\,\sigma\cdot\mathbf{k}=\left[\begin{array}{cc}
{\cal V}_0 & k_x+ ik_y \\ k_x - ik_y & {\cal V}_0 \end{array} \right]\ .
\label{AI:eq:2}
\end{equation}
Here, ${\cal V}_0$ is a uniform external potential. We will also adapt
the following notation: $k_\pm = k_x \pm i k_y$. The energy dispersions
are simply  $\varepsilon(k)=\beta\hbar v_F\,k=\beta\hbar v_F\sqrt{k_x^2+k_y^2}$ and gapless. In addition, $\beta$ is the electron-hole
parity index\,\footnote{The quantity $\beta=\pm 1$ is often called \textit{pseudo-spin} because of its formal resemblance to the spin index
in a spinor wavefunction.}, so that $\beta=1$ for electrons and $\beta=-1$ for holes.

The  goal of  this Section is to compare the electronic and
transport properties of  gapped graphene modeled by finite electron
effective  mass with those obtained  in the case of massless Dirac
fermions. There have been many studies\,\cite{bib:AI:substrate,bib:AI:gapstun}, where the mass term is added to the Dirac Hamiltonian of infinite
graphene. An energy gap may appear as a result of  a number of
physical reasons, such as  by a boron nitride substrate. However, the
most interesting cases are when the gap becomes \textit{tunable} and
may be varied throughout the experiment, resulting in practical
applications. We will focus on the so-called \textit{electron-photon
dressed states}, which result from the interaction between
Dirac electrons in graphene and circularly-polarized photons.
The first complete quantum description of such a system was
presented in Ref.\,\cite{bib:AI:Kibis}, and the transport
properties were discussed  in Ref.\,\cite{bib:AI:kibis2}.
The quantum descriptions for both graphene\,\cite{dh-ref-2.11}
and three-dimensional topological insulators were further
developed in Ref.\,\cite{bib:AI:mine2}.

\section{Tunneling in Gapped Graphene}

The so-called \textit{Klein paradox} is related to complete
unimpeded transmission of Dirac fermions through square potential
barriers of arbitrary height and width in the case of head-on
collision. It has been demonstrated that certain aspects of the
Klein paradox may also  be  observed in bilayer
graphene\,\cite{bib:AI:Katsnelson}, carbon nanotubes, topological insulators\,\cite{bib:AI:mine2} and zigzag nanoribbons\,\cite{dh-ref-2.11}.
The trademark of the Klein paradox exists even in the case of
electron-photon dressed states for massive electrons\,\cite{dh-ref-2.11}.

\subsection{Klein Paradox for the Square Barrier Tunneling}

\begin{figure}[ht!]
\centering
\includegraphics[width=0.85\textwidth]{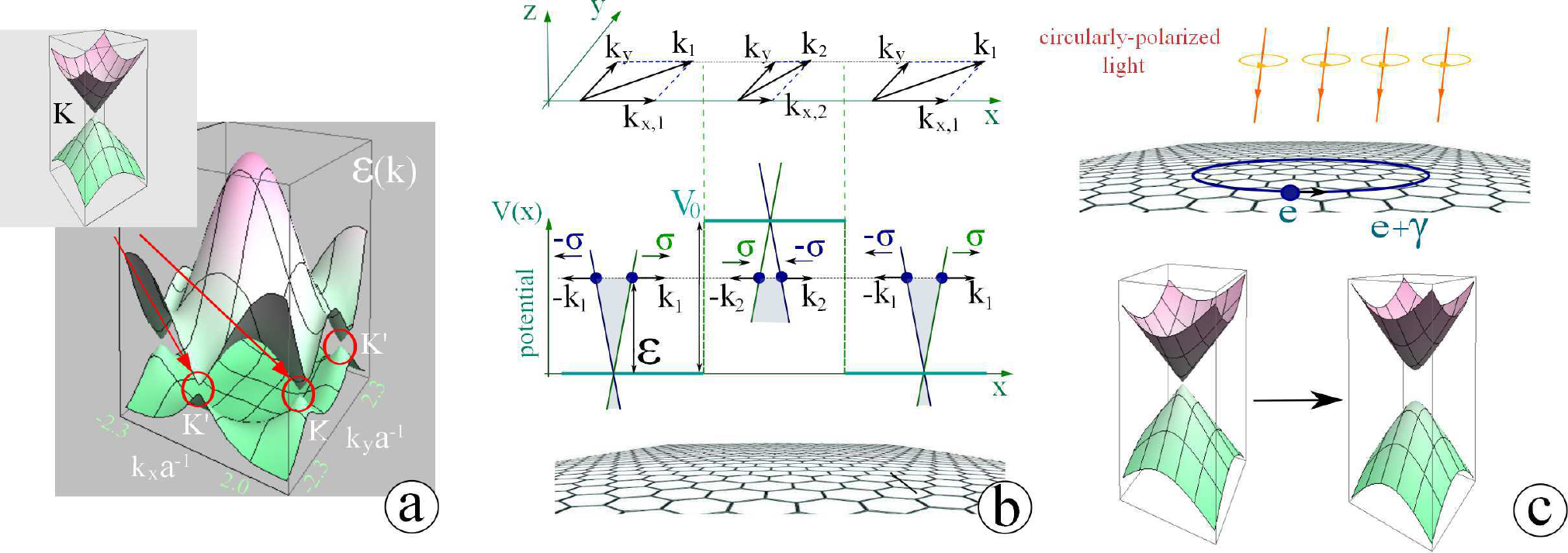}
\caption{Schematics of Dirac cone in graphene, square
barrier transmission and electron-photon dressed states.
(a) shows the energy bands $\varepsilon(k)$ as well as the
Dirac cone as a good approximation in the vicinity of
certain points in the ${\bf k}$-plane. (b) illustrates the
square barrier tunneling with conserved pseudo-spin, and
introduces all the notations used in the discussion of  tunneling.
(c) features the electron-photon dressed states, which appear as a
result of Dirac electrons interacting with circularly-polarized
photons. It also demonstrates schematically that the energy dispersion
of the electron dressed states has a gap, proportional to the intensity
of the light.} \label{AI-fig-1}
\end{figure}
In order to demonstrate Klein tunneling and investigate its
unique properties, we will consider a sharp
potential barrier (like a $p-n-p$ junction), given by $V(x)=V_0\left[\theta(x)-\theta(x-w) \right]$ and infinite
in the $y-$direction specified by the Heaviside step
function $\theta(x)$. Here, we will use the notations introduced in Fig.\,\ref{AI-fig-1}(b).

Klein tunneling may be explained based on a specific form of the Dirac
fermion wavefunction as well as a special type of \textit{chiral}
symmetry of the Dirac Hamiltonian. The eigenvalue wavefunctions
obtained from Eq.\,(\ref{AI:eq:1}) are $\varepsilon(k)=\hbar v_F\,k$ and

\begin{equation}
\Psi_{\beta}(k_x,\,k_y)=\frac{1}{\sqrt{2}}\left[ \begin{array}{c}
1 \\ \beta \texttt{e}^{i(k_x x+k_y y)}
\end{array}\right]\ .
\label{AI:eqn:3}
\end{equation}
Electrons are said to be chiral if their wavefunctions are eigenstates
of the chirality operator $\hat{h}=\sigma\cdot\hat{\mathbf{p}}/(2p)$,
where $\sigma=\{\sigma_x,\,\sigma_y\}$ is the Pauli vector consisting of
the Pauli matrices and $\hat{\mathbf{p}}=\{\hat{p}_x,\,\hat{p}_y\}$
is the electron momentum operator in graphene layers. Electrons become
chiral in graphene due to the fact that the chirality operator is
proportional to the Dirac Hamiltonian which automatically makes
\textit{chirality} a good quantum number. One can easily verify
that the wavefunction in Eq.\,(\ref{AI:eqn:3}) satisfies the chirality property.

As an example, we consider the situation of a very high potential
barrier,  i.e., $V_0 \gg\varepsilon$, which would not allow any finite
transmission amplitude possible for a conventional Schr\"odinger
particle. For Dirac electrons, however, we have

\begin{equation}
T(k_{x,1},\,k_{x,2})=\frac{\cos^2\phi}{\cos^2{(k_{x,2}w)\,\cos^2\phi+\sin^2 (k_{x,2}w)}}\ ,
\end{equation}
where $\phi$ is the angle which ${\bf k}$ makes with the $x$-axis.
Incidentally, for the case of head-on collision with $\phi=0$, we
find complete unimpeded tunneling with $T=1$, which is a direct
consequence of a special form of the electron wavefunction,
resulting from the Dirac cone energy dispersion.

\section{Electron-Photon Dressed States with Gap}

The peaks of transmission  mainly belong to two different species.
Namely, the \textit{Klein paradox} for head-on collision where
$\phi=0$ and the so-called ``transmission resonances'' correspond
to specific values of the electron longitudinal momenta in the barrier region.

It was shown recently\,\cite{bib:AI:Kibis,bib:AI:kibis2} that when
Dirac electrons in a single graphene layer  interact with an intense
circularly polarized light beam, electron states will be \textit{dressed}
by photons. Here, we  investigate the transmission properties
of such dressed electrons for the case of single layer graphene.

We begin with the electron-photon interaction Hamiltonian

\begin{equation}
\label{AI:eq:4}
\hat{\mathcal{H}}=v_F\,\sigma \cdot \left({\hat{\mathbf{p}}-e\,\mathbf{A}_{circ}}\right) \ ,
\end{equation}
where the vector potential for circularly polarized light of
frequency $\omega_0$ may be expressed as

\begin{equation}
\label{AI:eq:5}
\mathbf{A}_{circ}=\sqrt{\frac{\hbar}{\epsilon_0\,\omega_0{\cal V}}} \left({ \mathbf{e}_+ \hat{a} + \mathbf{e}_- \hat{a}^{\dag} }\right)
\end{equation}
in terms of photon creation and destruction operators
$\hat{a}^{\dag}$ and $\hat{a}^{\dag}$, respectively.
Here, ${\cal V}$ is the mode volume of an optical field. In
order to study the complete electron-photon interacting
system, we must add the field energy term $\hbar\omega_0\,
\hat{a}^{\dagger}\hat{a}$ to the Hamiltonian Eq.\,(\ref{AI:eq:4}).

In terms of the energy dispersion $\varepsilon(k)=\pm\sqrt{(\hbar v_F k)^2+\Delta^2}$, our system is formally similar to the eigenvalue
equations for the case of the effective mass $\sigma_3$ Dirac Hamiltonian:
\begin{equation}
\label{AI:eq:6}
\hat{\mathcal{H}}=\hbar v_F\,\sigma \cdot \mathbf{k}+
V(x) \left[{\begin{array}{cc}
1 & 0 \\ 0 & 1
\end{array} }\right]+\Delta\,\sigma_{3}\ ,
\end{equation}
where $\sigma_{3}$ is a Pauli matrix and $V(x)$ is a
one-dimensional potential. The electron dispersion and transmission
properties for  both a single and multiple square potential
barrier have been studied\,\cite{bib:AI:Barb-review,bib:AI:Barb2}
for monolayer and bilayer graphene. It was also shown that a
one-dimensional periodic array of potential barriers  leads to
multiple  Dirac points\,\cite{bib:AI:Barb-review}. Several published works
have introduced an effective mass term into the Dirac Hamiltonian for infinite
graphene, which may be justified based on  different physical
reasons\,\cite{bib:AI:gapstun,bib:AI:pnp}. For example, it has been
shown\,\cite{bib:AI:substrate} that an energy band gap in graphene can be
created by boron nitride substrate, resulting in a
finite electron effective mass. However, we would like to emphasize that the
analogy between the Hamiltonian Eq.\,(\ref{AI:eq:6}) and that for
irradiated graphene is not complete since the former requires
$\Delta<0$. Although this difference does not result in
any modification of the energy dispersion term containing
$\Delta^2$, it certainly changes the corresponding wavefunction.
Additionally, there have been a number of studies using laser
radiation  on single layer\,\cite{bib:AI:r1} and bilayer\,\cite{bib:AI:Luis1} graphene as well as graphene nanoribbons\,\cite{bib:AI:c1} reporting
the gap opening as a result of strong electron-photon interaction.
From these aspects, it seems that topological insulators as well as
gapped graphene may have potential device applications where spin plays a role.
In the presence of a strong radiation field, we have the dressed-state wavefunction
\begin{equation}
\Phi_{dr}(k)= \left[{ \begin{array}{c}
\mathcal{C}_1(k) \\
\beta\,\mathcal{C}_2(k)\,\texttt{e}^{i\phi}
\end{array} }\right]\ ,
\end{equation}
where $\mathcal{C}_1(k)\neq\mathcal{C}_2 (k)$ and are given by
\begin{eqnarray}
\mathcal{C}^{\pm}_1(k)=\frac{1}{\sqrt{2(1+\gamma^2) \mp 2\gamma\sqrt{1+\gamma^2}}}\ ,\\
\mathcal{C}^{\pm}_2(k)=\pm
\frac{\sqrt{1+\gamma^2}\mp\gamma}{\sqrt{2(1+\gamma^2) \mp
2\gamma\sqrt{1+\gamma^2}}}\ .
\end{eqnarray}
Here, $\gamma=\Delta/(\hbar v_Fk)$.
Consequently, the chiral symmetry is broken for electron dressed states.
The interaction between Dirac electrons in graphene and circularly
polarized light has been considered in the classical limit  in
Ref.\,\cite{bib:AI:aoki_oki}. In this limit, a gap  in the Dirac cone opens
up due to nonlinear effects. The dressed-state wavefunction has the chirality
\begin{equation}
\hat{h}\Phi_{dr}(k)=\frac{1}{2}\,\frac{\sigma
\cdot\hat{\mathbf{p}}}{p}\left[{\begin{array}{c}
\mathcal{C}_1(k)\\
\beta\,\mathcal{C}_2(k)\,\texttt{e}^{i \phi}
\end{array}}\right]=\frac{1}{2}\left[{\begin{array}{c}
\beta\,\mathcal{C}_2(k) \\
\mathcal{C}_1(k)\,\texttt{e}^{i \phi}
\end{array}}\right]\  .
\end{equation}
As it appears, non-chirality of the dressed electron states becomes
significant if the electron-photon interaction (the leading $\gamma$ term) is
increased. This affects electron tunneling and transport
properties. We now turn to an investigation of the transmission of
electron states through a potential barrier when graphene is
irradiated with  circularly polarized light.

\par
Summarizing the description of the electron dressed states, we would like to estimate 
the actual values of the energy gap in graphene and describe its dependence on the power 
and the frequency of the applied laser beam, which obeys the equation:
\begin{equation}
\Delta = \sqrt{ \mathbb{W}_0^2 + \left( \hbar \omega_0 \right)^2 } - \hbar \omega_0 \, , 
\end{equation}
where $\mathbb{W}_0  = 2 v_F \mathcal{E}_0 / \omega_0$, so that the gap is proportional to the square of amplitude of the electric 
field $\mathcal{E}_0$ for $\mathcal{E}_0 \longrightarrow 0$. The amplitude of the electromagnetic field is determined by the laser  power
and its wavelength $\mathcal{E}_0 \backsimeq P^{1/2} \lambda^{-1}$. Consequently, the energy gap in graphene depends on both the power of the
applied laser and its wavelength. The results could be summarized in the following table:

\begin{table}
\begin{tabular}{| l || c | c | c | r |}
\hline
\hline
        &   $\lambda=10^{-7}$ m & $\lambda=10^{-6}$ m & $\lambda=10^{-5}$ m  \\
\hline

P = 0.1 W    &    0.027         &             0.274    &          2.71        \\

\hline

P = 1 W      &    0.264         &             2.738    &           25.03      \\

\hline

P= 100 W     &    27.39         &             249.14   &           709.213     \\

\hline
\hline
\end{tabular}
\caption{Energy gap in graphene, corresponding to the different values of the laser power $P$ and its wavelength $\lambda$. 
The gap obviously increases with the increasing power, similar dependence on $\lambda$ is observed, since the field amplitude  is inversely proportional to the corresponding frequency. The energy gap dependence is approximately a linear function of the laser power for short wavelength.}
\end{table}

\par

\subsection{Tunneling and Optical Properties of Electron Dressed States}

We consider a square potential barrier given by
$V(x)=V_0[{\theta(x)-\theta(x-W_0)}]$. The current component is $j_x=\Phi^{\dag}\sigma_x\Phi$, thus we only require the wave-function
continuity at the potential boundaries. There are two specific
simplifications to consider here - a nearly-head-on collision
$k_{x,i} \ll k_{y,i}$ and the case of a high potential barrier
when $\varepsilon \ll V_0$. We should also mention another relevant study\,\cite{bib:AI:gomes}, investigating the tunneling of Dirac electrons
with a finite effective mass through a similar potential barrier.

For a nearly-head-on collision with $k_y\ll k_{x,1}\ll k_{x,2}$ for
high potential as well as  infinite graphene ($\Delta\to 0$),
the transmission coefficient has the following simplified form
\begin{equation}
T=1-\sin^2(k_{x,2} W_0)\,\left(\theta^2-2\beta\theta\phi+\phi^2\right)\ ,
\end{equation}
where we assume $V_0\gg\varepsilon$, $\theta\ll\phi\ll 1$ and
$\beta=\pm 1$.
\begin{figure}[ht!]
\centering
\includegraphics[width=0.95\textwidth]{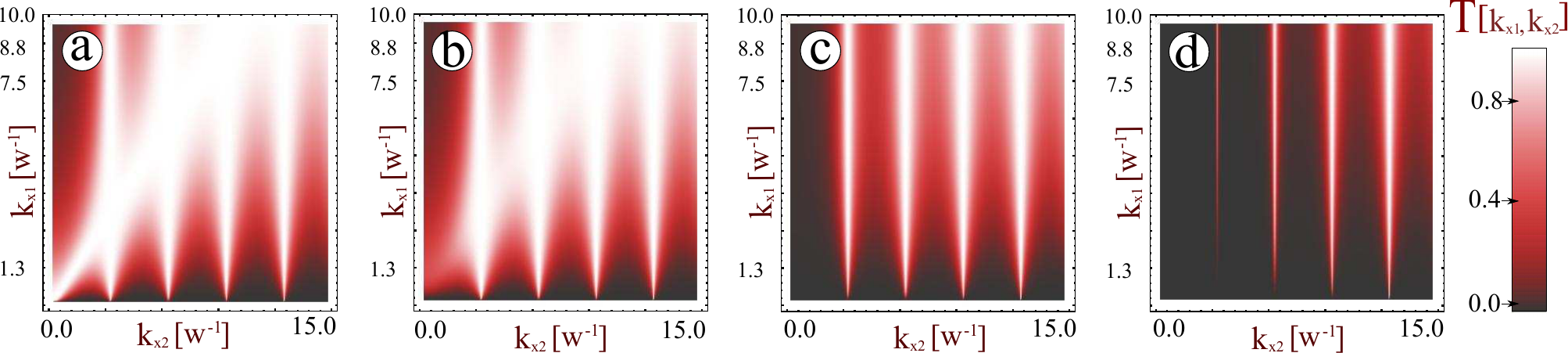}
\caption{Density plots of the transmission probability $T$ for
electron-photon dressed states as a function of the electron
longitudinal momenta $k_{x,1}$ and $k_{x,2}$ in both \textit{barrier}
$V(x)=V_0$ and \textit{no-barrier} $V(x)=0$ regions,
respectively. The electron momenta are given in the units of the inverse barrier width $w$. Here, plots (a), (b), (c) and (d)
correspond to $\Delta/V_0=0$, $0.001$, $0.008$ and $0.015$, respectively.}
\label{AI-fig-2}
\end{figure}

Figure\ \ref{AI-fig-2} presents the calculated transmission probability
$T$ as a function of the longitudinal momenta $k_{x,1}$ (in front the
barrier) and $k_{x,2}$ (in the barrier region). We find from
Fig.\,\ref{AI-fig-2} that the intensity of the transmission peaks in
(b)-(d) are gradually distorted with increasing gap compared to
infinite graphene in (a). The diagonal $k_{x,1}=k_{x,2}$ corresponds
to the absence of a potential barrier and should yield a complete
transmission for $\Delta=0$ as seen in (a). However, for a finite $\Delta$, the
requirement $(\sqrt{(\varepsilon-V_0)^2-\Delta^2}>\hbar v_F\,k_y)$
must be satisfied, which makes the diagonal transmission
incomplete, e.g., missing diagonal for small $k_{x,1}$ and $k_{x,2}$ in (b), due to
the occurrence of an energy gap. As $\Delta$ is further increased in (c) and (d),
this diagonal distortion becomes more and more severe, which is accompanied by strongly reduced intensity of transmission peaks at small $k_{x,2}$.
\begin{figure}[ht!]
\centering
\includegraphics[width=0.49\textwidth]{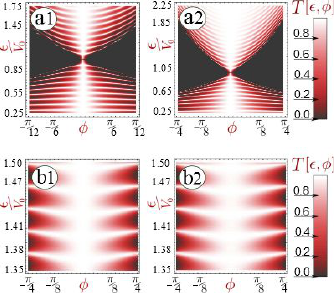}
\caption{Density plots of $T$ for electron-photon dressed states
as a function of the incoming electron energy $\varepsilon$ and
the angle of incidence $\phi$. The upper panel displays $T$ under
$\Delta=0$ in (a1) as well as its blowout views in (a2)-(a4).
The lower panel is associated with various induced gaps
$\Delta/V_0=0.01$, $0.05$, $0.07$ and $0.1$ in (b1)-(b4).
The significant difference in tunneling behavior at $\phi=0$
can be seen with various $\Delta$ values.}
\label{AI-fig-3}
\end{figure}

Figure\ \ref{AI-fig-3} displays the effect of the energy gap on $T$ in
terms of incoming particle energy $\varepsilon$ and angle of incidence
$\phi$. From the upper panel of Fig.\,\ref{AI-fig-3}(a1)-(a4), we
see the Klein paradox as well as other resonant tunneling peaks in
$T$ for regular infinite graphene with $\Delta=0$. The dark ``pockets''
on both sides of $\varepsilon=V_0$ demonstrate zero transmission
for the case of $\vert\varepsilon-V_0\vert\ll\varepsilon$, which
results in imaginary longitudinal momentum $k_{x,2}$ for most of
the incident angles and produces a fully attenuated wavefunction.
\begin{figure}[ht!]
\centering
\includegraphics[width=0.8\textwidth]{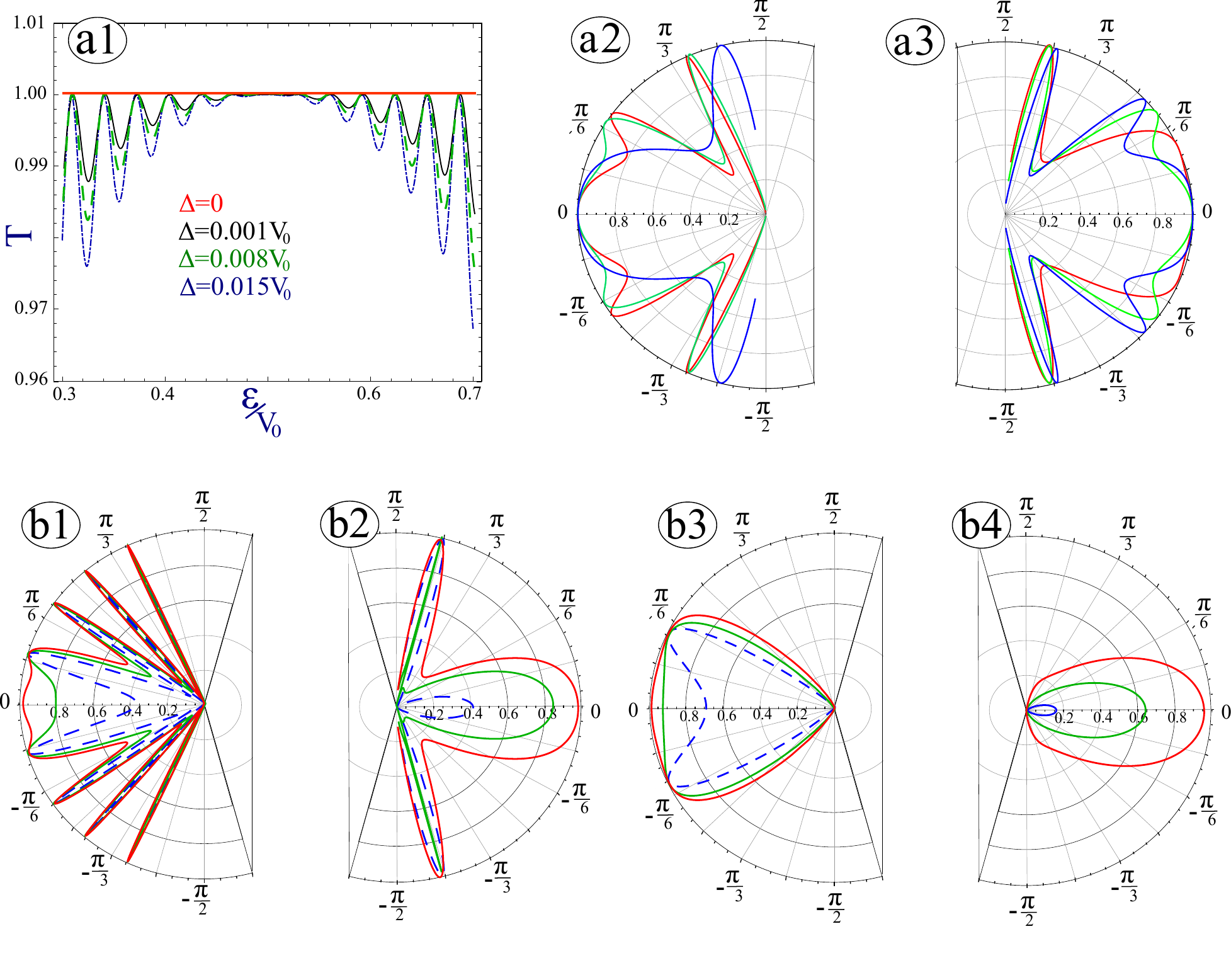}
\caption{The upper panel presents $T$ as functions of $\varepsilon$ in
(a1) and $\phi$ in (a2) and (a3). In (a1), $T$ with $\phi=0$ for
various $\Delta$ values are exhibited with $\Delta=0$ (red),
$\Delta/V_0=0.001$ (black), $\Delta/V_0=0.008$ (green)
and $\Delta/V_0=0.015$ (blue). In (a2) and (a3), the electron-photon
interaction is excluded and $\varepsilon=V_0/6$. Additionally, we assume
in (a2) $V_0=100$\,meV and various $W_0$ values with $W_0=5$\,nm (red),
$W_0=10$\,nm (green) and $W_0=20$\,nm (blue). Plot (a3) shows
$T$ with $W_0=20$\,nm for $V_0=150$\,meV (red), $V_0=200$\,meV
(green) and $V_0=250$\,meV (blue). The lower panel presents dressed-state $\phi$ dependence of $T$ with $\varepsilon=V_0/6$ for $V_0=0.5$\,eV and
$W_0=100$\,nm in (b1); $V_0=2$\,eV and $W_0=100$\,nm in (b2);
$V_0=4$\,eV and $W_0=150$\,nm in (b3); and $V_0=5$\,eV and
$W_0=150$\,nm in (b4). Moreover, we set in (b1)-(b4)
$\Delta=0.099$\,eV (blue), $\Delta=0.050$\,eV (green) and
$\Delta=0.010$\,eV (red).}
\label{AI-fig-4}
\end{figure}

Figure\ \ref{AI-fig-4} shows $T$ as a function of $\varepsilon$ in
(a1) and of $\phi$ in all other plots. From (a1), we clearly see that
dressing destroys the Klein paradox for head-on collision with $\phi=0$.
Additionally, in the absence of electron-photon interaction, we display
the effects on $T$ for various values of $W_0$ in (a2) and $V_0$ in (a3).
The resonant peaks are found to be shifted toward other incoming angles
in both (a2) and (a3) and this effect becomes stronger for small incident
angles. The effect of dressed electron states, on the other hand, is
demonstrated in each plot in (b1)-(b4) for various values of $\Delta$,
as well as in individual plots of (b1)-(b4) for different barrier
heights $V_0$ and widths $W_0$.

It is very helpful to compare the results obtained here with bilayer
graphene having quadratic dispersion. For bilayer graphene, its lowest
energy states are described by the Hamiltonian\,\cite{bib:AI:Katsnelson}
\begin{equation}
\hat{\mathcal{H}}_{blg}=\frac{\hbar^2}{2 m_b}\left(k_-^{2}\sigma_++k_+^{2}\sigma_-\right)\ ,
\end{equation}
where $m_b$ is the effective mass of electrons in the barrier region. In this
case, the longitudinal wave vector component in the barrier region is
given by $k_{x,2}=\beta^{\prime}\sqrt{2m_b\beta(\varepsilon-V_0)-k_y^2}$
with $\beta$, $\beta^{\prime}=\pm 1$.

An evanescent wave having a decay rate $\kappa_b$ may coexist with a
propagating wave having a wave vector $k_{x,2}$ such that
$k_y^2+k_{x,2}^2=k_y^2-\kappa_b^2=2 m_b\beta(\varepsilon-V_0)$.
This implies that the evanescent modes should be taken into account
simultaneously. The Klein paradox persists in bilayer graphene
for chiral but massive particles. However, one finds a complete reflection,
instead of a complete transmission, in this case.

In the simplest approximation to include only the two nearest subbands,
our model is formally similar to the so-called $\sigma_3$
Hamiltonian used for describing the particles in a single layer of
graphene with parabolic energy dispersion (non-zero effective mass).
By including more than two independent pairs, this effect is expected
to be weaker since perfect transmission will occur only for the two
wavefunction terms. However, this effect is insensitive to the barrier width, and
therefore, may be considered as reminiscent of the Klein paradox.

\par

In spite of the fact that a significant number of papers on electron
tunneling in the presence of the electron-photon interaction have already 
been published, this topic is receiving a lot of attention at the present 
time. Potential barrier transmission is studied using the Floquet approach
\cite{bib:AI:Bis,bib:AI:Hungary}.
It is appears that the valley of the \textit{massive} Dirac electrons could become an important factor for the square barrier tunneling. Valleytronics uses the valley degree of freedom as a carrier of information similarly to the way spintronics uses electron spin \cite{Moldovan}

 The effect of periodic potential is similar 
to the combination of the uniform magnetic and electric fields \cite{bib:AI:HofsConf}.
Also we would like to mention Ref.\cite{bib:AI:Mod}, where the transport properties were
investigated in the framework of  unitary-transformation scheme together with the 
non-equilibrium Green's function formalism. Similar case of Chiral tunneling modulated 
by a time-periodic potential on the surface states of a topological insulator was addressed
in \cite{bib:AI:Nature2}. 

\par

In conclusion, we would like to mention \textit{topological insulators}. Due to their band structure and a nontrivial topological
order, topological insulators are insulating in the bulk, but support
gapless conducting surface states\,\cite{bib:AI:HasanTI}. The surface states
are represented by a spin-polarized Dirac cone, creating an analogy with
graphene. However, there is a peculiar property of the topological
insulator, i.e., a so-called geometrical gap in the energy dispersion.
This gap appears if the sample is finite in the $z$-direction, which has
not been observed in conventional insulators. The electron dressed
states in three dimensional topological insulators have been obtained and their
 tunneling properties have been discussed\,\cite{bib:AI:mine2,bib:AI:mySPIE}.

\section{Enhanced Mobility of Hot Dirac Electrons in Nanoribbons}
\label{dh-sec1}

Early studies\,\cite{dh-ref-1.16,dh-ref-1.11} on transport in graphene
nanoribbons (GNRs) were restricted to the low-field limit\,\cite{dh-ref-1.01},
where a linearized Boltzmann equation with a relaxation-time approximation
was solved. Recently, the non-equilibrium distribution of electrons is
calculated by solving the Boltzmann equation beyond the relaxation-time approximation for nonlinear transport in semiconducting
GNRs\,\cite{dh-ref-1.02}. Enhanced mobility from field-heated electrons
in high energy states is predicted. An anomalous enhancement in the
line-edge roughness scattering under high fields is found with decreasing
roughness correlation length due to the population of high-energy states by field-heated electrons\,\cite{dh-ref-1.0}.

\subsection{Nonlinear Boltzmann Theory}
\label{dh-sec1-1}

We limit ourselves to single subband transport with low electron densities,
moderate temperatures, ionized impurities and line-edge roughness\,\cite{dh-ref-1.16,dh-ref-1.18,dh-ref-1.19}.
Consequently, negligible influences from electron-electron\,\cite{dh-ref-1.20},
optical phonon\,\cite{dh-ref-1.19}, inter-valley and volume-distributed impurity scattering\,\cite{dh-ref-1.16} are expected.
The discrete energy dispersion for Armchair-nanoribbons (ANRs) is written as\,\cite{dh-ref-1.4}
\begin{equation}
\label{dh-eq-1.1}
\varepsilon_j = \hbar v_F\times\Bigg\{{\begin{array}{ll}
k_j &,\ \ \ \ \ \ \ \ \mbox{metallic}\\
\sqrt{k^2_j +\left({\pi/3W}\right)^2} &,\ \ \ \ \ \ \ \ \mbox{semiconducting}
\end{array}}\ .
\end{equation}
The discrete wave numbers are $k_j=[{j-(N+1)/2}]\delta k$ with $j=1,\,2,\,\ldots,\,N$ for a large odd integer $N$,
and $\delta k=2\,k_{\rm max}/(N-1)$ is mesh spacing.
The central point is $j=M=(N+1)/2$ for the minimum of the energy.
$W=(\mathcal{N}+1)\,a_0/2$ is the GNR width, $a_0=2.6$\,\AA\
the size of the graphene unit cell, and $\mathcal{N}$ the number
of carbon atoms across GNRs. From Eq.\,(\ref{dh-eq-1.1}),
we get $v_F$ for the group velocity $v_j$ for metallic nanoribbons,
while for semiconducting ANRs,
it is $v_j=v_F\left({\hbar\nu_{F}k_j/\varepsilon_j}\right)$.
We know the band will be filled up to $\lvert{k_j}\rvert=k_F$
at zero temperature ($T=0$\,K) with the Fermi
wave number and Fermi energy given, respectively, by $k_F=\pi n_{1D}/2$ and $\varepsilon_F=\varepsilon(k_F)$. For a chosen $T$ and chemical
potential $\mu_0$, the linear density in ANRs is $\displaystyle{n_{1D}=\delta k/\pi\,\sum_{j=1}^N\,[{\exp{((\varepsilon_j-\mu_0)/\mathrm{k}_B T)}+1}]^{-1}}$.

We assume that the wavefunction $\Psi_j(x,\,y)$ corresponding to Eq.\,(\ref{dh-eq-1.1}) satisfies hard-wall boundary
conditions\,\cite{dh-ref-1.21} $\Psi_j(0,\,y)=\Psi_j(W,\,y)=0$. This can
be fulfilled by selecting the wavefunction as a mixture of ones at $\displaystyle{\mathbf{K}=\left({\frac{2\pi}{3a_0},
\frac{2\pi}{\sqrt{3}a_0}}\right)}$ and
$\displaystyle{\mathbf{K}^\prime=\left({-\frac{2\pi}{3a_0},
\frac{2\pi}{\sqrt{3}a_0}}\right)}$ points as\,\cite{dh-ref-1.4}
\begin{equation}
\Psi_j(x,\,y)=\frac{1}{\sqrt{2}
}\left[{\psi_j (x,\,y)-\psi^\prime_j(x,\,y)}\right]\ ,
\label{dh-eq-1.2}
\end{equation}
\begin{equation}
\left\{
\begin{array}{ll}
\psi_{j}(x,\,y)=\sqrt{\frac{1}{2LW}}\,e^{ik_jy}\,\left[{\begin{array}{c} 1\\ e^{i\phi_{j}}\end{array}}\right]\,e^{i(2\pi/3a_0-\kappa)x}
&\ \ \ \ \mbox{at $\mathbf{K}$ point}\\
\psi^\prime_{j}(x,\,y)=\sqrt{\frac{1}{2LW}}\,e^{ik_jy}\,\left[{\begin{array}{c} 1\\ -e^{-i\phi_{j}}\end{array}}\right]\,e^{-i(2\pi/3a_0-\kappa)x}
&\ \ \ \ \mbox{at $\mathbf{K}^\prime$ point}
\end{array}\right.\ .
\label{dh-eq-1.2+}
\end{equation}
Here, $L$ is the ribbon length. For semiconducting ANRs
$\kappa=\pi/3W\ll 2\pi/3a_0$ is the quantum of the transverse wave vector, and
$\phi_j=\tan^{-1}\left({k_j/\kappa}\right)$ the phase separation
between the two graphene sublattices. For a metallic-type  ribbon, we
set $\kappa=0$ and the phase assumes only $\pm\pi/2$ values.

With the wavefunction in Eq.\,(\ref{dh-eq-1.2}) and neglecting inter-valley
scattering\,\footnote{Such inter-valley scattering would require momentum transfer comparable with the distance between $\mathbf{K}$ and $\mathbf{K}^\prime$ points.} one may calculate the scattering from any potential $V(x,\,y)$.
The impurity and phonon induced inter-valley scattering is neglected because the relevant phonon energy for momentum transfer
is large at low temperatures and the effective scattering cross section of both
volume and surface impurities is suppressed for large value of $|\mathbf{K}-\mathbf{K}^\prime|$.
Therefore, the interaction matrix elements become
\begin{equation}
V_{i,\,j}=\int_{0}^{W} dx\int^{\infty}_{-\infty} dy\,\Psi^\ast_i(x,\,y)\,V(x,\,y)\,\Psi_j(x,\,y)
=\frac{1}{2}\,\int_{0}^{W} dx\int^{\infty}_{-\infty} dy\,\left({\psi^\ast_i\,V\,\psi_j+\psi^{\prime\,\ast}_i\,
V\,\psi^\prime_j}\right)\ .
\label{scatter}
\end{equation}
We consider scattering potential made from the three contributions to be
$V=V^{\rm AL}+V^{\rm LER}+V^{\rm imp}$.

Since the longitudinal phonons induce higher deformation potential than the out-of-plane flexural ones, we neglect the flexural modes here. Under this approximation, the phonon-scattering potential is written as\,\cite{dh-ref-1.16,dh-ref-1.23}
\begin{equation}
\label{dh-eq-1.3}
V^{\rm AL}(y)=\sqrt{\frac{n^{\pm}\hbar}{2\rho LW\omega_{AL}}}\,
D_{AL}q_y\,e^{iq_yy}\ ,
\end{equation}
where
$n^{-} = \left[{\exp{(\hbar\omega_{AL}/\mathrm{k}_B T)-1}}\right]^{-1}$ and $n^{+}=1+n^{-}$ are the equilibrium phonon distributions,
$\omega_{AL}=c_s\,q_y$ the phonon frequency,
$D_{AL}\sim 16$\,eV the deformation potential,
$\rho\sim 7.6\times 10^{-8}$\,g/cm$^2$, and
$c_s\sim 2\times 10^6$\,cm/s the mass density and sound velocity.
Moreover, the momentum conservation gives $q_y=k_i-k_j$.

Elastic scattering is attributed to the roughness of the ribbon edges
and in-plane charged impurities. For the former, we assume the width
of the ribbon as $W(y)=W+\delta W (y)$ and the edge-roughness to
satisfy the Gaussian correlation function
$\langle\delta W(y)\,\delta W(y+\Delta y)\rangle
=\delta b^2\exp[-(\Delta y/\Lambda_0)^2]$
with $\delta b\sim 5$\,\AA\ being the amplitude and
$\Lambda_0\sim 50-200$\,\AA\ the correlation length.
The latter is related to impurities located at $(x_0,\,0,\,0)$
and distributed with a sheet density $n_{\rm 2D}$.
Each point impurity produces a scattering potential.
Since the momentum difference between two valleys is
very large, only short-range impurities can contribute.
Here, we neglect the short range scatterers thus making
the scattering matrix diagonal. Corresponding perturbations for roughness and impurity are
\begin{eqnarray}
\label{dh-eq-1.4}
V^{\rm LER}(y) & = & \frac{\delta W (y)}{3W^2}\,\pi\hbar v_F\ ,\\
\notag
\\
\label{dh-eq-1.5}
V^{\rm imp}(x,\,y) & = & \frac{e^2}{4\pi\epsilon_0\epsilon_r\sqrt{(x-x_0)^2 + y^2}}\ ,
\end{eqnarray}
where $\epsilon_r$ is the average host dielectric constant.
These potentials provide the net elastic scattering rate $\hbar/\tau_j$ through
\begin{gather}
\label{dh-eq-1.6}
\frac{1}{\tau_j} = \frac{1}{\tau^{imp}_j}+\frac{1}{\tau^{LER}_j}\ ,\\
\notag
\\
\label{dh-eq-1.7}
\left\{
\begin{array}{ll}
\left(\tau^{imp}_j\right)^{-1}=\gamma_0\,\left(\frac{v_{\rm F}}{|v_j|}\right)\,\left[1+\cos(2\phi_j)\right]\\
\left(\tau^{LER}_j\right)^{-1}=\gamma_1 \left({\frac{v_F}{|v_j|}}\right)\frac{1}{1+4k_j^2\Lambda_0^2}\,\left[1+\cos(2\phi_j)\right]
\end{array}\right.\ ,
\end{gather}
where $\gamma_0$ is the scattering rate of impurities at the Fermi edge
and $\displaystyle{\gamma_1=2\left(\frac{\pi v_F\delta b}{3W^2}\right)^2\frac{\Lambda_0}{v_F}}$ the scattering rate from edge
roughness. Here, we assume the impurities distribute within the layer.
Moreover, the scattering potential is screened by electrons. Since we limit ourselves to a single subband, the screening is just
a scalar Thomas-Fermi dielectric function\,\cite{dh-ref-1.16,dh-ref-1.26}
$\epsilon_{TF}(\lvert{k_{j^\prime}-k_j}\rvert)$, which is
calculated as $\displaystyle{\epsilon_{TF}\approx 1+\frac{e^2}{\pi^2\epsilon_0\epsilon_r\hbar v_F}}$
in the metallic limit ($2k_F W\gg 1$) with $\epsilon_r\approx 3.9$.
We apply the static screening to both impurity and phonon scattering
with a relative large damping rate and small Fermi energy.

The deviation from the Fermi distribution under a strong field
is described by the nonlinear Boltzmann equations\,\cite{dh-ref-1.18,dh-ref-1.19}
\begin{equation}
\frac{dg^\prime_j(t)}{dt}=b_j-\sum_{j^{\,\prime}\neq M}\,a^\prime_{j,\,j^{\,\prime}}(t)\,g^\prime_{j^{\,\prime}}(t)\ ,
\label{dh-eq-1.8}
\end{equation}
where $g^\prime_j(t) = g_j(t)-g_M (t)$ is the reduced form of the non-equilibrium
part\,\footnote{A distinct line must be drawn between equilibrium distribution function $f_j^{(0)}$ in absence of applied electric field and
stationary solution of the transport equation $\lim\limits_{t\rightarrow 0}\,f_j(t)$.} of the distribution. This reduced form ensures conservation
of particle numbers, i.e.  $\displaystyle{\sum_{j=1}^N g_j(t)=0}$.
Equation (\ref{dh-eq-1.8}) was used for investigating the dynamics in quantum wires\,\cite{dh-ref-1.18}
and quantum-dot superlattices\,\cite{dh-ref-1.19}.
In Eq.\,(\ref{dh-eq-1.8}), we introduced notation for the matrix elements
$a^\prime_{j,\,j^{\,\prime}}(t)=a_{j,\,j^{\,\prime}}(t)-a_{j,\,M}(t)$
via its components
\begin{eqnarray}
a_{j,\,j^{\,\prime}}(t)&=&\delta_{j,\,j^{\,\prime}}\left[{\cal W}_j+{\cal W}_j^g(t)+
\frac{1-\delta_{j,\,(N+1)/2}}{2\tau_j}\right]-
\delta_{j+j^{\,\prime},\,N+1}\left[\frac{1-\delta_{j,\,(N+1)/2}}{2\tau_j}\right]\\
\notag
&-& {\cal W}_{j,\,j^{\,\prime}}-\frac{e{\cal F}_0}{2 \hbar\delta
k} \left({\delta_{j,j^\prime-1}-\delta_{j,j^\prime+1}}\right) \ .
\label{dh-eq-1.9}
\end{eqnarray}
The total inelastic rate is $\displaystyle{\mathcal{W}_j=
\frac{1}{\tau_{j}^{AL}}=\sum_{j^\prime}\,\mathcal{W}_{j,j^\prime}}$
with the scattering matrix
\begin{equation}
\label{dh-eq-1.10}
{\cal W}_{j,\,j^{\,\prime}}=\frac{L}{2\pi}\,\delta k \sum_\pm\,{\cal W}^\pm_{j,j^\prime}\,\left({n_{j,j^\prime}+f^{\pm}_{j^\prime}}\right)\ ,
\end{equation}
\begin{equation}
W^{\pm}_{j,j^\prime}=
\theta(\pm\varepsilon_{j^{\,\prime}}\mp\varepsilon_j)\,\left[
\frac{D_{AL}^2|\varepsilon_{j^{\,\prime}}-\varepsilon_j|}{2\hbar^2c_s^3
\rho L W \epsilon^2_{\rm TF}(|k_{j^{\,\prime}}-k_j|)}\right]\,
\left[1+\cos(\phi_{j^{\,\prime}}-\phi_j)\right]\ .
\end{equation}
Here, we employ the notations $f^{-}_j=f^{(0)}_j$, $f^{+}_j=1-f^{(0)}_j$, $n_{j,j^\prime}
=N_0(|\varepsilon_{j^{\,\prime}}-\varepsilon_j|/
\hbar)$, and $N_0(\omega_q)=[\exp(\hbar\omega_q/k_{\rm B}T)-1]^{-1}$ is the Bose function for phonons.
Moreover, the nonlinear phonon scattering rate ${\cal W}_j^g(t)$,
associated with the heating of electrons, is
\begin{eqnarray}
{\cal W}_j^g(t)=\frac{L}{2\pi}\,\delta k\,
\sum_{j^{\,\prime}\neq M}\,g^\prime_{j^{\,\prime}}(t) \left[{\cal W}^{+}_{j,j^\prime}
-{\cal W}^{-}_{j,j^\prime}-\left({\cal W}^{+}_{j,M}-{\cal W}^{-}_{j,M}\right)\right]\ .
\label{dh-eq-1.11}
\end{eqnarray}
Once $g^\prime_j(t)$ is obtained from Eq.\,(\ref{dh-eq-1.8}), the
drift velocity $v_c(t)$ is found from
\begin{equation}
\label{dh-eq-1.12}
v_{\rm c}(t)=\left[{\sum\limits_{j=1}^{N}\,f^{(0)}_j}\right]^{-1}\times
\left\{
\begin{array}{ll}
{\sum\limits_{j\neq M}\,\left(v_j -v_M\right)\,g^\prime_j(t)} & ,\ \ \ \  \mbox{semiconducting}\\
2v_F\sum\limits_{j=1}^{M}\,g^\prime_j(t) & ,\ \ \ \  \mbox{metallic}
\end{array}
\right.\ .
\end{equation}
One knows that the equilibrium part of the distribution does not
contribute to the drift velocity. The steady-state drift velocity
$v_{\rm d}$ is obtained from $v_{\rm c}(t)$ in the limit $t \to \infty$. Furthermore, the steady-state current is given by $I=en_{1D}v_{ \rm
d}$. The differential mobility is defined as
$\mu_e=\partial v_{\rm d}/\partial\mathcal{F}_0$.

\subsection{Numerical Results for Enhanced Hot-Electron Mobility}
\label{dh-sec1-2}
\begin{figure}[ht!]
\centering
\includegraphics[width=0.95\textwidth]{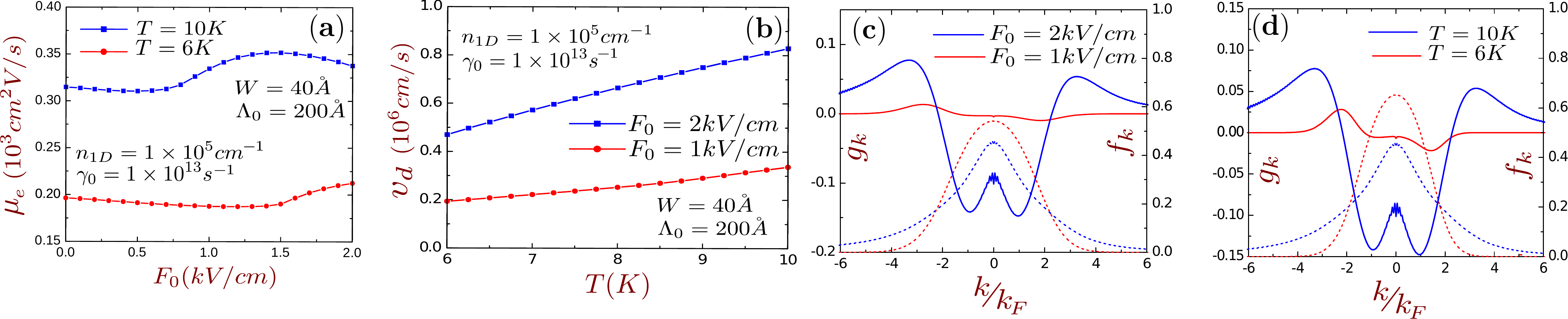}
\caption{(a) Calculated electron mobilities $\mu_{\rm e}$ as a function of applied
electric field ${\cal F}_0$ at $T=10$\,K (solid squares on blue curve) and $T=6$\,K
(solid circles on red curve); (b) Electron drift velocities $v_{\rm d}$ as a function of temperature  $T$ at ${\cal F}_0=2$\,kV/cm (blue curve) and ${\cal F}_0=1$\,kV/cm (red curve); (c) The non-equilibrium part of ($g_k$, left-hand scaled solid curves) and total ($f_k$, right-hand scaled dashed curves) electron distribution functions at $T=10$\,K as functions of electron wave number
$k$ along the ribbon with ${\cal F}_0=2$\,kV/cm (blue curves) and ${\cal F}_0=1$\,kV/cm (red curves); (d) $g_k$ (left-hand scaled solid curves) and $f_k$ (right-hand scaled dashed curves) with ${\cal F}_0=2$\,kV/cm as functions of $k$ at $T=10$\,K (blue curves) and $T=6$\,K (red curves). The other parameters are
indicated directly in (a) and (b).}
\label{dh-fig-1.1}
\end{figure}

Figure\ \ref{dh-fig-1.1}(a) displays the mobilities $\mu_{\rm e}$ as
a function of ${\cal F}_0$ at $T=10$\,K
and $T=6$\,K. We see from Fig.\,\ref{dh-fig-1.1}(a) a strong
${\cal F}_0$-dependence at a lower value of ${\cal F}_0$ and
at higher temperature $T$. The observed feature
comes from the electron-phonon scattering
rate ${\cal W}_j^g(t)$ in Eq.\,(\ref{dh-eq-1.11}). We find
a lower threshold field ${\cal F}^\ast$ is needed for entering
into a nonlinear regime (${\cal F}>{\cal F}^\ast$) because of
enhanced phonon scattering at $T=10$\,K.
${\cal F}^\ast$ strongly depends on
$T$, $n_{\rm 1D}$, $\gamma_0$ and $\Lambda_0$. As ${\cal F}_0\to 0$,
$\mu_{\rm e}$ is larger at $T=10$\,K than at $T=6$\,K since there
exists an additional thermal population of high-energy states with a
high group velocity. The initial reduction of $\mu_{\rm e}$ is connected
to the increased frictional force with ${\cal F}_0$ from phonon scattering.
We can see $\mu_{\rm e}$ is independent of ${\cal F}_0$ below
$0.75$\,kV/cm (linear regime) at $T=10$\,K. However, $\mu_{\rm e}$ goes up
significantly above $0.75$\,kV/cm (nonlinear regime). Finally,
$\mu_{\rm e}$ drops with ${\cal F}_0$ beyond $1.5$\,kV/cm (heating regime),
giving rise to a saturation of the drift velocity. From Eq.\,(\ref{dh-eq-1.1}) we understand the group velocity $|v_j|$ increases with $|k_j|$.
However, this increase becomes slower as approaching to $v_F$.
Physically, the rise of $\mu_{\rm e}$ with ${\cal F}_0$ in the
nonlinear regime comes from the initially-heated electrons in high
energy states with a larger group velocity, while the successive drop
of $\mu_{\rm e}$ in the heating regime connects to the combination of the
upper limit $v_j\leq v_F$ and the significantly enhanced phonon scattering. In Fig.\,\ref{dh-fig-1.1}(b), $v_{\rm d}$ is presented as a function of $T$ at
${\cal F}_0=2$\,kV/cm and ${\cal F}_0=1$\,kV/cm. $\mu_{\rm e}$ increases
with $T$ monotonically in both cases, proving that the scattering is not dominated
by phonons but by impurities and line-edge roughness. Different features
in the increase of $\mu_{\rm e}$ are attributed to the linear and nonlinear
regimes. For ${\cal F}_0=2$\,kV/cm in the  nonlinear regime, $v_{\rm d}$ (or
$\mu_{\rm e}$) increases with $T$ sub-linearly, while $v_{\rm d}$ goes
up super-linearly with $T$ in the linear regime at ${\cal F}_0
=1$\,kV/cm. Different $T$ dependence in $\mu_{\rm e}$ directly connects to
$g_j$, displayed in Figs.\,\ref{dh-fig-1.1}(c) and \ref{dh-fig-1.1}(d). $g_j$
at $T=10$\,K, as well as the total distribution function $f_j$,
are exhibited in Fig.\,\ref{dh-fig-1.1}(c) as functions of $k_j$
at ${\cal F}_0=2$\,kV/cm and ${\cal F}_0=1$\,kV/cm. The electron
heating is found from Fig.\,\ref{dh-fig-1.1}(c) at ${\cal F}_0=2$\,kV/cm
by moving thermally driven electrons from low to high-energy states
with heat resulting from the work done by a frictional
force\,\cite{dh-ref-1.23} due to phonon scattering.
At ${\cal F}_0=1$\,kV/cm, we find electrons swept by elastic scattering
from the right Fermi edge to the left one in the linear regime.
Figure\ \ref{dh-fig-1.1}(d) demonstrates a comparison between $g_j$
and $f_j$ at $T=10$\,K and $T=6$\,K under ${\cal F}_0=2$\,kV/cm, showing that phonon scattering
is important at $T=10$\,K, while the elastic scattering of electrons
dominates at $T=6$\,K, which agrees with the observation in Fig.\,\ref{dh-fig-1.1}(c).
\begin{figure}[ht!]
\centering
\includegraphics[width=0.9\textwidth]{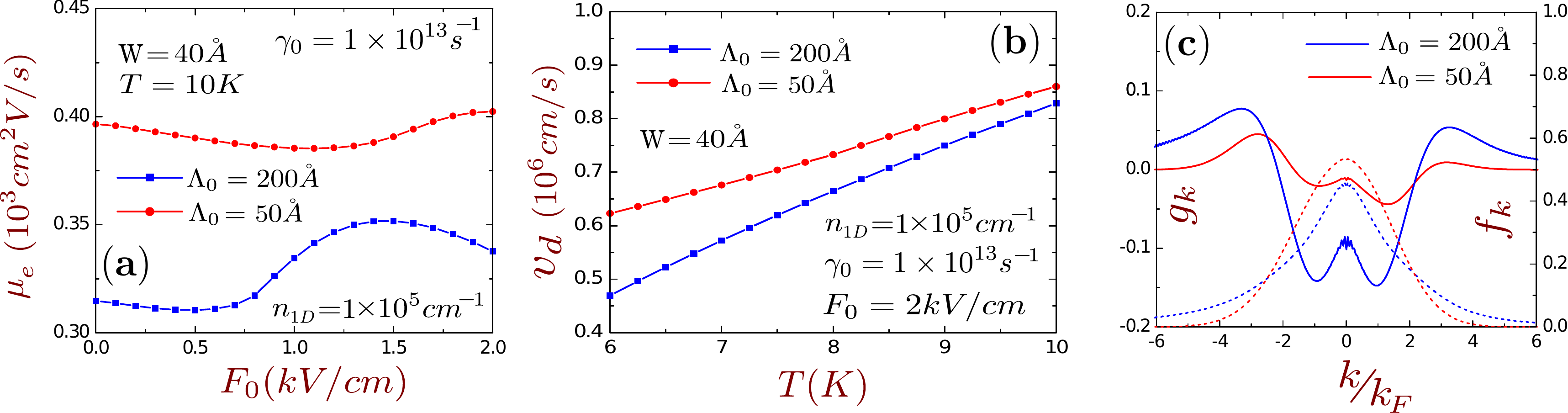}
\caption{(a) $\mu_{\rm e}$ as a function of ${\cal F}_0$ at $T=10$\,K with
$\Lambda_0=200$\,\AA\ (solid squares on blue curve) and $\Lambda_0=50$\,\AA\
(solid circles on red curve); (b) $v_{\rm d}$ as a function of
$T$ with ${\cal F}_0=2$\,kV/cm for $\Lambda_0=200$\,\AA\ (solid squares
on blue curve) and $\Lambda_0=50$\,\AA\ (solid circles on red curve);
(c) $g_k$ (left-hand scaled solid curves) and $f_k$ (right-hand scaled
dashed curves) as a function of $k$. Here, the cases with $\Lambda_0=
200$\,\AA\ and $\Lambda_0=50$\,\AA\ are represented by blue and red curves, respectively. The other parameters are indicated in (a) and (b).}
\label{dh-fig-1.2}
\end{figure}

The correlation-length effect for the line-edge roughness (LER) is
shown in Figs.\,\ref{dh-fig-1.2}(a)-(c) by setting $W=50$\,\AA\ and
varying $\Lambda_0$ from $200$\,\AA\ through $50$\,\AA. We know from Eq.\,(\ref{dh-eq-1.7}) that the LER scattering may either decrease or
increase with $\Lambda_0$, depending on $|k_j|\ll 1/2\Lambda_0$
or $|k_j|\gg 1/2\Lambda_0$. For $n_{\rm 1D}=1.0\times 10^5$\,cm$^{-1}$,
we know $|k_j|\ll 1/2\Lambda_0$ is met in the low-field limit
($|k_j|\sim k_{\rm F}$), while we have $|k_j|\gg 1/2\Lambda_0$ for the
high field limit due to electron heating. Consequently, we see from Fig.\,\ref{dh-fig-1.2}(a) $\mu_{\rm e}$ increases as
${\cal F}_0\to 0$ when $\Lambda_0$ is $50$\,\AA\ in the low-field
regime. However, ${\cal F}^\ast$ increases in the high-field regime
for $\Lambda_0=50$\,\AA. This is accompanied by a reduction in the enhancement
of $\mu_{\rm e}$. The anomalous feature with $\Lambda_0$ gives a profound
impact on the $T$-dependence of $\mu_{\rm e}$ as shown in  Fig.\,\ref{dh-fig-1.2}(b), where the rising rate of $\mu_{\rm e}$
with $T$ in the high-field regime is much smaller with
$\Lambda_0=50$\,\AA\ than for $\Lambda_0=200$\,\AA.
Additionally, $g_j$ in Fig.\,\ref{dh-fig-1.2}(c) exhibits an anomalous cooling
behavior, i.e. with a smaller spreading of $f_j$ in the $k_j$ space, in the high-field regime as $\Lambda_0$ drops to $50$\,\AA.

\section{Magnetoplasmons in Gapped Graphene}

Electronic and transport proprieties of Dirac electrons in the presence
of a uniform perpendicular magnetic field have been reported in a few studies\,\cite{bib:AI:MFReviewGoerbig,bib:AI:MFReviewKrotov,bib:AI:Gus1}.
The wavefunction, obtained in\,\cite{bib:AI:MFAndo1}, has a number of
novel and intriguing properties. Additionally, electron-electron Coulomb
interaction is found to have effects on both the quasiparticle effective mass
and collective excitations \,\cite{bib:AI:GGGfactor}. Using the same method as
for the case of circularly-polarized light radiation in Eq.\,(\ref{AI:eq:1}), we further take into account the vector potential
for a uniform perpendicular magnetic field, giving
\begin{eqnarray}
\label{AI:eq:7}
\hat{\mathcal{H}} &=& v_F\,\sigma\cdot\left({\mathbf{p}-e\mathbf{A}}\right)\ , \\
\nonumber
{\bf B} &=& \nabla\times{\bf A}=\{0,\,0,\,B\}\ .
\end{eqnarray}
In the case of 2DEG, on the other hand, the Hamiltonian contains the effective electron mass $m^\ast$:
\begin{equation}
\label{AI:eq:8}
\hat{\mathcal{H}}_{2DEG}=\frac{1}{2m^\ast}\,\left(\mathbf{p}-e\mathbf{A}\right)^2\\
\end{equation}
with similar canonical momentum substitution.
\begin{figure}[ht!]
\centering
\includegraphics[width=0.85\textwidth]{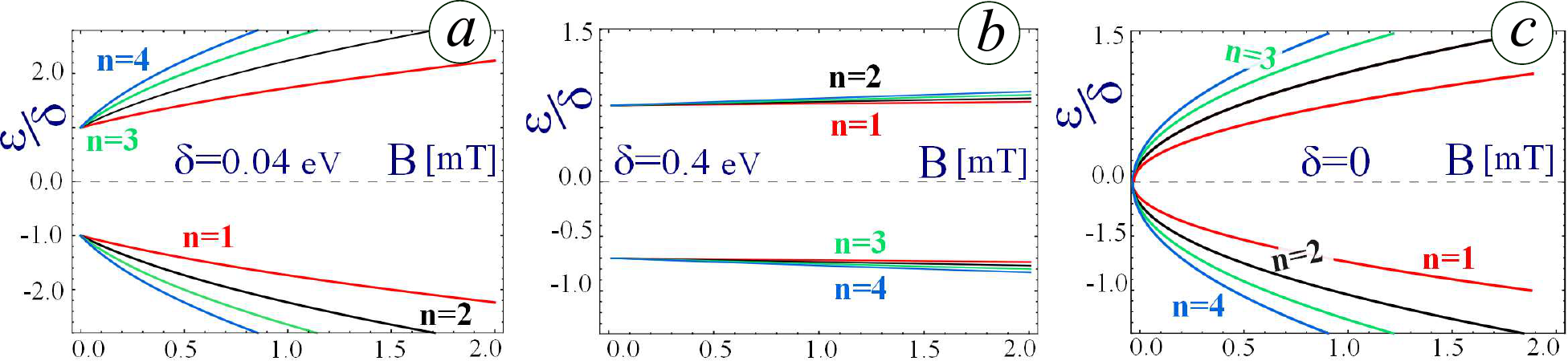}
\caption{Dirac electron energy eigenvalues (Landau levels) as a
function of the perpendicular magnetic field $B$.
The first four levels for both electrons and holes are displayed.
Plots (a)-(c) correspond to  various energy gaps, i.e.,  $\Delta=0.04$\,eV, $\Delta=0.4$\,eV and $\Delta=0$, respectively.}
\label{AIfig5}
\end{figure}

Although the energy levels are equidistant for 2DEG with ($n=0,1,2,\cdots$),
$\varepsilon_n=\omega_c(n+1/2)$ and $\omega_c$ the cyclotron frequency,
in graphene the difference between  two consecutive energy levels decreases
with the index $n$ of the level as $\backsimeq 1/\sqrt{n}$, i.e., $\varepsilon_n=(\hbar v_F/\ell_B)\sqrt{2n}$. Here   $\ell_B=\sqrt{\hbar/(eB)}$.
Also,  $\omega_0=\hbar v_F/\ell_B$ will be used as a unit of frequency
for both the 2DEG and graphene. These eigenenergies apparently depend
on the applied magnetic field $B$ and are presented in Fig.\,\ref{AIfig5}.
The $B$ dependence becomes the strongest for the case of zero gap in
(c) and negligible with a large gap in (b).

In order to obtain  the one-loop polarization function $\Pi^{0}(q,\,\omega)$,
we perform a summation over all possible transitions between the states
on both sides of the Fermi energy, so that both occupied and unoccupied
states will be included. The   coefficients in the  summations, representing
the weight of each term, are called the \textit{oscillator strengths} or form factors. Obviously for two well-separated eigenstates
with $\vert n_1-n_2\vert\gg 1$, their coefficients become infinitesimal,
which in leading order may be expressed as a factorial function
$\displaystyle{\frac{(\vert n_1-n_2\vert)!}{(n_1+n_2)!}}$. The general
definition of the form factor is the wavefunction overlap, i.e.,
\begin{equation}
{\cal F}\left(\frac{q^2\ell_B^2}{2}\right)=\vert\langle n_2,\beta_2\,\vert\texttt{e}^{i\bf{q}\cdot\bf{r}}\vert n_1,\beta_1\rangle\vert^2\ ,
\end{equation}
where $\beta_1,\,\beta_2=\pm 1$. We will directly write down the expressions
for form factors of both 2DEG and graphene. In the case of 2DEG, the
form factor is calculated as
\begin{eqnarray}
{\cal F}_{n_1,n_2}^{2DEG}(q) &=& \texttt{e}^{-\xi}\,\xi^{\vert n_1-n_2\vert}\,\frac{n_<!}{n_>!}\,\left[\mathcal{L}_{n_<-1}^{\vert n_1-n_2\vert}(\xi)\right]^2\ ,\\
\nonumber
\xi &=& \frac{q^2\ell_B^2}{2}\ ,
\end{eqnarray}
where $n_<$ is the lesser of two integers $n_1$ and $n_2$ while $n_>$ is the greater. In the case of \textit{gapped} graphene with $\Delta>0$,
the form factor becomes
\begin{equation}
{\cal F}^{n_1,n_2}_{\beta_1,\beta_2}(q)=\texttt{e}^{-\xi}\,\xi^{\vert n_1-n_2\vert}\left[\left(1+\frac{\beta_1\beta_2
\Delta^2}{|\varepsilon_{n_1}\varepsilon_{n_2}|}\right)
\mathcal{P}_1+\mathcal{P}_2\right]\ ,
\end{equation}
\[
{\mathcal P}_1=\frac{n_<!}{n_>!}\left[\mathcal{L}_{n_<}^{\vert n_1-n_2\vert}(\xi)\right]^2+(1-\delta_{0n_<})\,\frac{(n_<-1)!}{(n_>- 1)!}\left[\mathcal{L}_{n_<-1}^{\vert n_1-n_2\vert}(\xi)\right]^2\ ,\\
\]
\[
\mathcal{P}_2=\frac{4\beta_1\beta_2 v_F^2}{\ell_B^2|\varepsilon_{n_1}\varepsilon_{n_2}|}\,\frac{n_{<}!}{(n_>-1)!}\,\mathcal{L}_{n_<}^{\vert n_1-n_2\vert}(\xi)\,\mathcal{L}_{n_<-1}^{\vert n_1-n_2\vert}(\xi)\ ,\\
\]
\[
\vert\varepsilon_n\vert=\sqrt{\frac{2\hbar^2 v_F^2}{\ell_B^2}\,n+\Delta^2}\ .
\]
For the case of zero energy gap, $\Delta \to 0$, one may easily obtain the
standard expression for the form factor in graphene, similar to the
results in Ref.\,\,\cite{bib:AI:Rafael0,bib:AI:Rafael1,bib:AI:Pyat2}.
The function $\Pi^{0}(q,\,\omega)$ contains two principal parts, i.e., the , vacuum polarization corresponding to  \textit{interband} transitions in
\textit{undoped} graphene, as well as another part involving the summation over all the occupied Landau level ($N_F$ is the highest occupied Landau level)),
taking into consideration the \textit{intraband} transitions.

Our goal is to explore both incoherent (particle-hole modes) and
coherent (plasmons) excitations in the system. The imaginary part of $\Pi^{0}(q,\,\omega)$ defines the structure factor $\mathcal{D}(q,\,\omega)$, i.e.,
\begin{equation}
\mathcal{D} (q,\,\omega)=-\frac{1}{\pi}\,{\rm Im}\Pi^{0}(q,\,\omega)\ .
\end{equation}
The particle-hole mode region is obtained from the condition
${\rm Im}\Pi^{0}(q,\,\omega)>0$. Additionally, the plasmons become
\textit{damping free} only if they stays away from the particle-hole
mode region, or equivalently, $\mathcal{D}(q,\,\omega)=0$.
On the other hand, the plasmon frequencies are defined by the zeros
of the dielectric function $\epsilon(q,\,\omega)=1-v_c(q)\,\Pi^{0}(q,\,\omega)$.
Here, $\displaystyle{v_c(q)=\frac{e^2}{2\epsilon_0\epsilon_b\,q}}$ is the Fourier-transformed unscreened Coulomb potential and $\epsilon_b$ is
the background dielectric constant.  These plasmons  may also be
obtained from the peak of the renormalized
polarization $\Pi^{\rm RPA}(q,\,\omega)$ for interacting electrons
in the random-phase approximation (RPA), giving
\begin{equation}
\Pi^{\rm RPA}(q,\,\omega)=\frac{\Pi^{0}(q,\,\omega)}{1- v_c(q)\,\Pi^{0}(q,\,\omega)}\ .
\end{equation}
We   note that a metal-insulator transition may also occur under
a magnetic field\,\cite{bib:AI:MFGapGorbar}, similar to the effect of circularly-polarized light, along with other effects\,\cite{bib:AI:MFFertig1,bib:AI:MFFertig2,bib:AI:MFAndo2}.

\subsection{Single-Particle Excitations and Magnetoplasmons}

We now present and discuss our numerical results of both $\Pi^0(q,\,\omega)$ and $\Pi^{\rm RPA}(q,\,\omega)$ for various values of chemical potentials $\mu$
(or equivalently, number of occupied Landau levels $N_F$) and energy gap $\Delta$. The disorder broadening is set to be $\eta=0.05\,\hbar v_F/\ell_B$ for all our numerical calculations. One of the major effects found in our numerical results
is the magnetoplasmon for various energy gaps, compared with those for
gapless graphene (Dirac cone). We also compare $\Pi^{RPA}(q,\,\omega)$ for interacting electrons with non-interacting $\Pi^{0}(q,\,\omega)$ with various interaction
parameters $\displaystyle{r_s=\frac{2m^\ast e^2}{\epsilon_0\epsilon_b\,\hbar^2k_F}}$ for 2DEG and $\displaystyle{r_s=\frac{e^2}{\epsilon_0\epsilon_b\,\hbar v_F}}$ for graphene as well.
\begin{figure}[ht!]
\centering
\includegraphics[width=0.95\textwidth]{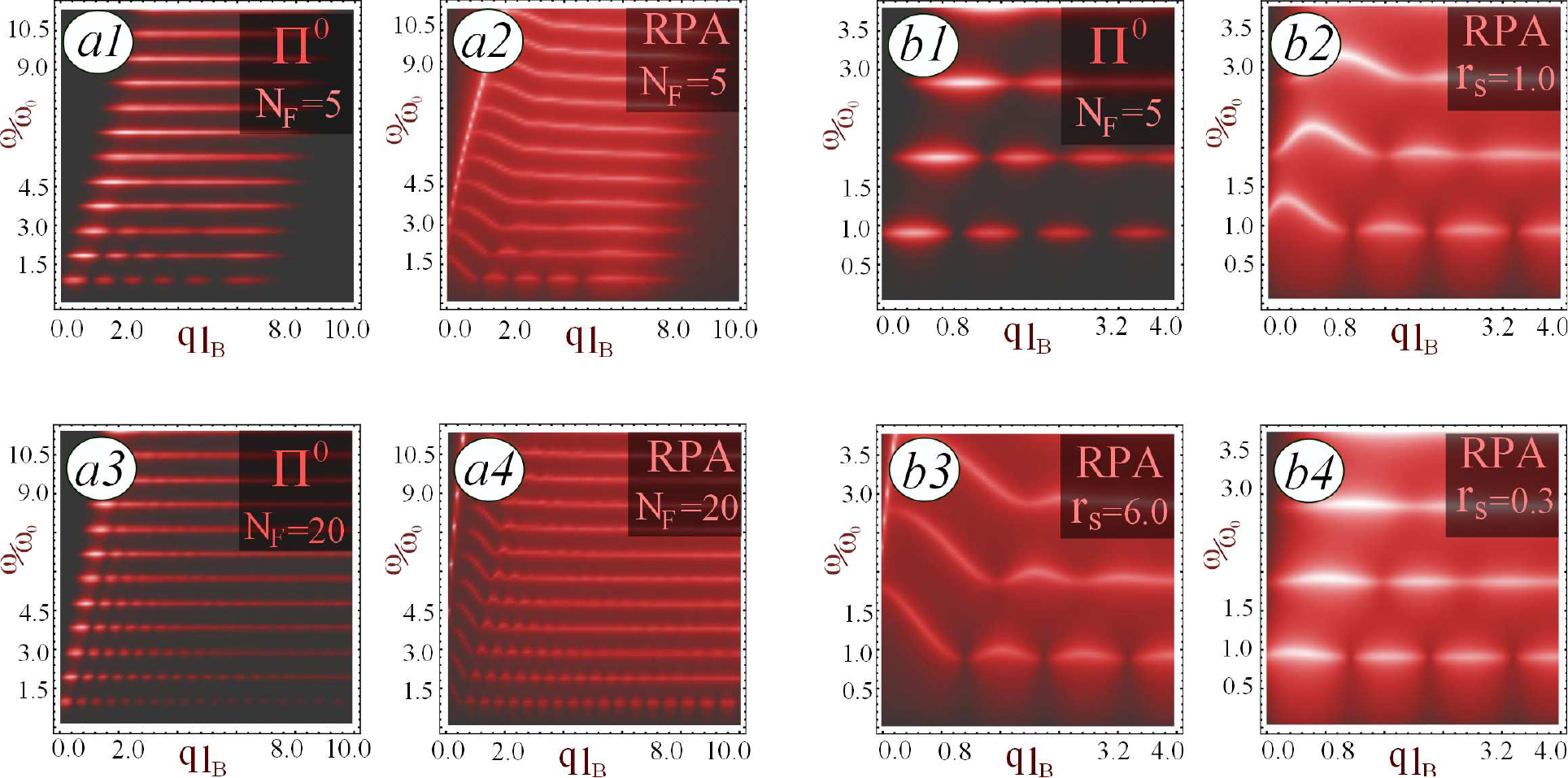}
\caption{Density plots of ${\rm Im}[\Pi^{0}(q,\,\omega)]$ and ${\rm Im}[\Pi^{RPA}(q,\,\omega)]$ for \textbf{2DEG}. The left panel (a1)-(a4)
presents non-interacting polarizations in (a1) and (a3) as well as the
renormalized polarizations in (a2) and (a4) for various values of $N_F$.
Plots (a1) and (a2) compare the effects due to  $N_F=5$, whereas plots
(a3) and (a4) are for the case of $N_F=20$. The left panel (b1)-(b4)
demonstrates ${\rm Im}[\Pi^{0}(q,\,\omega)]$ in (b1) and ${\rm Im}[\Pi^{\rm RPA}(q,\,\omega)]$ in (b2)-(b4) for $N_F=5$. Each plot in (b2)-(b4) corresponds to chosen interaction parameters $r_s=6.0$, $1.0$ and $0.3$, respectively.}
\label{AI-fig-6}
\end{figure}

It is interesting to compare the results in the presence of  an external magnetic field with those at zero magnetic field\,\cite{dh-ref-2.13,dh-ref-2.14,bib:AI:mySPIE,GG:PLAS:SOI-2}
for both 2DEG and gapped graphene. We emphasize  that the boundaries of the particle-hole continuum are drastically different for the 2DEG and graphene.
Whereas the edges of the particle-hole excitation region of 2DEG are parabolic, as seen from Fig.\,\ref{AI-fig-6}), these boundaries change to either straight lines in graphene (zero gap)
or to be slightly modified by the gap. Consequently, the plasmon region defined by $\mathcal{D}(q,\,\omega)=0$ in graphene is significantly increased even for a small energy gap $\Delta$, in comparison with that for 2DEG.
This demonstrates a large influence by a finite magnetic field on both the particle-hole excitation spectrum and the boundaries of the particle-hole mode regions at the same time.
\begin{figure}
\centering
\includegraphics[width=0.6\textwidth]{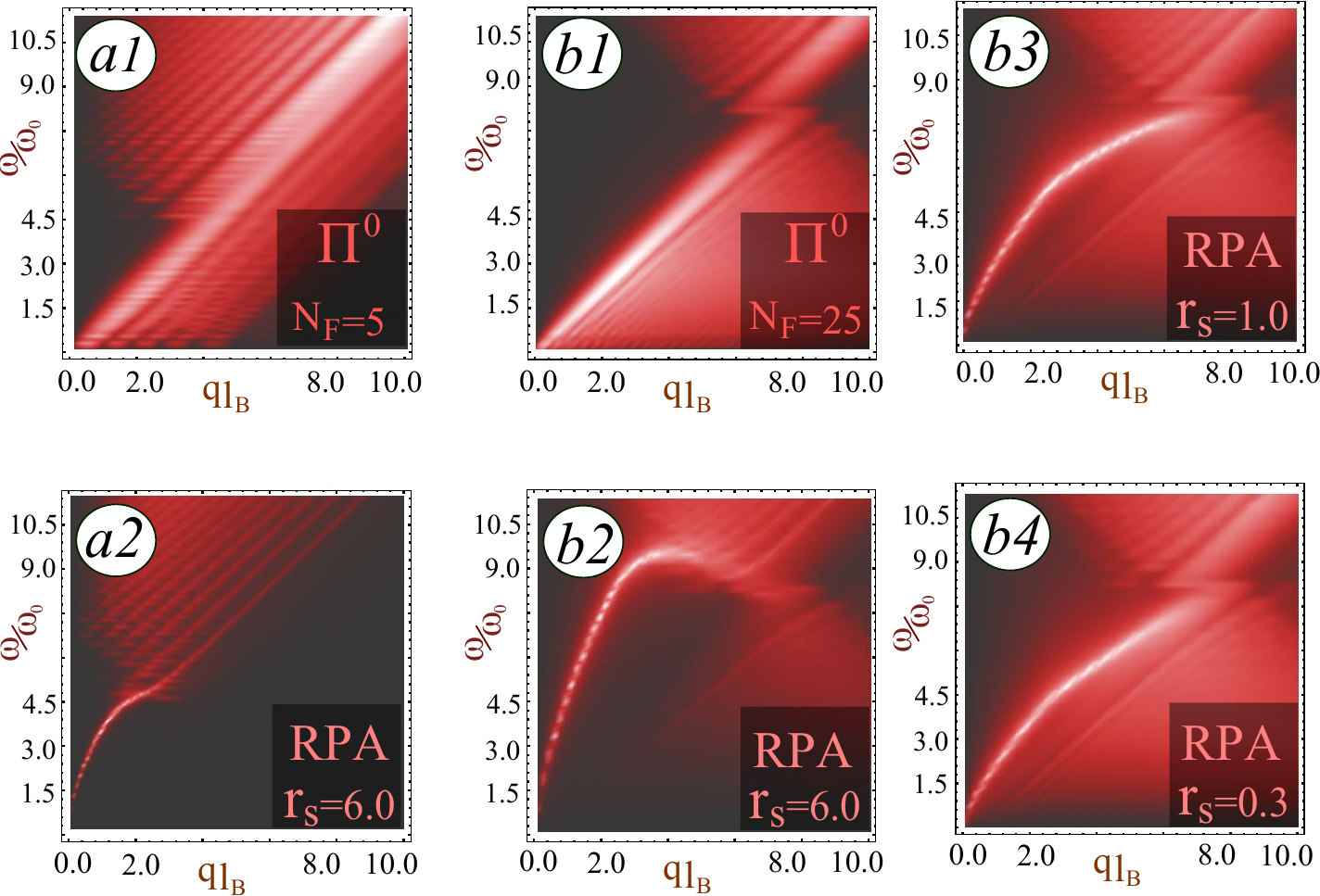}
\caption{Density plots of ${\rm Im}[\Pi^{0}(q,\,\omega)]$ and ${\rm Im}[\Pi^{\rm RPA}(q,\,\omega)]$ for \textbf{gapless graphene}.
The left panel (a1) and (a2) represents the non-interacting and RPA renormalized polarization functions with $N_F=5$. Plot (b1) corresponds to the non-interacting polarization function with $N_F=25$,
while plots (b2)-(b4) are associated with the RPA renormalized ones. Additionally, plots (b2)-(b4) present results  for $r_s=6.0$, $1.0$ and $0.3$, respectively.}
\label{AI-fig-7}
\end{figure}

Comparing Fig.\,\ref{AI-fig-6} with Fig.\,\ref{AI-fig-7}, we conclude that there exist two competing types of particle-hole excitation spectral behavior.
These are the horizontal lines with embedded islands for 2DEG and the inclined
lines for graphene. All the lines are well aligned with the boundaries of particle-hole mode regions at zero-field, including parabolic curves for 2DEG, straight lines for gapless graphene or nearly-straight lines for gapped graphene.
We note that the modulation for horizontal lines in 2DEG becomes   strongest,
in comparison with the inclined line in graphene. This difference is attributed to the fact that the Landau levels in graphene are \textit{not equidistant}
in contrast to those in 2DEG. Additionally, horizontal modulation prevails for a small broadening parameter  $\eta$, which has been verified experimentally for the range of $\eta$ within $0.05$--$0.5\,\bar v_F/\ell_B$.
\begin{figure}[ht!]
\centering
\includegraphics[width=0.8\textwidth]{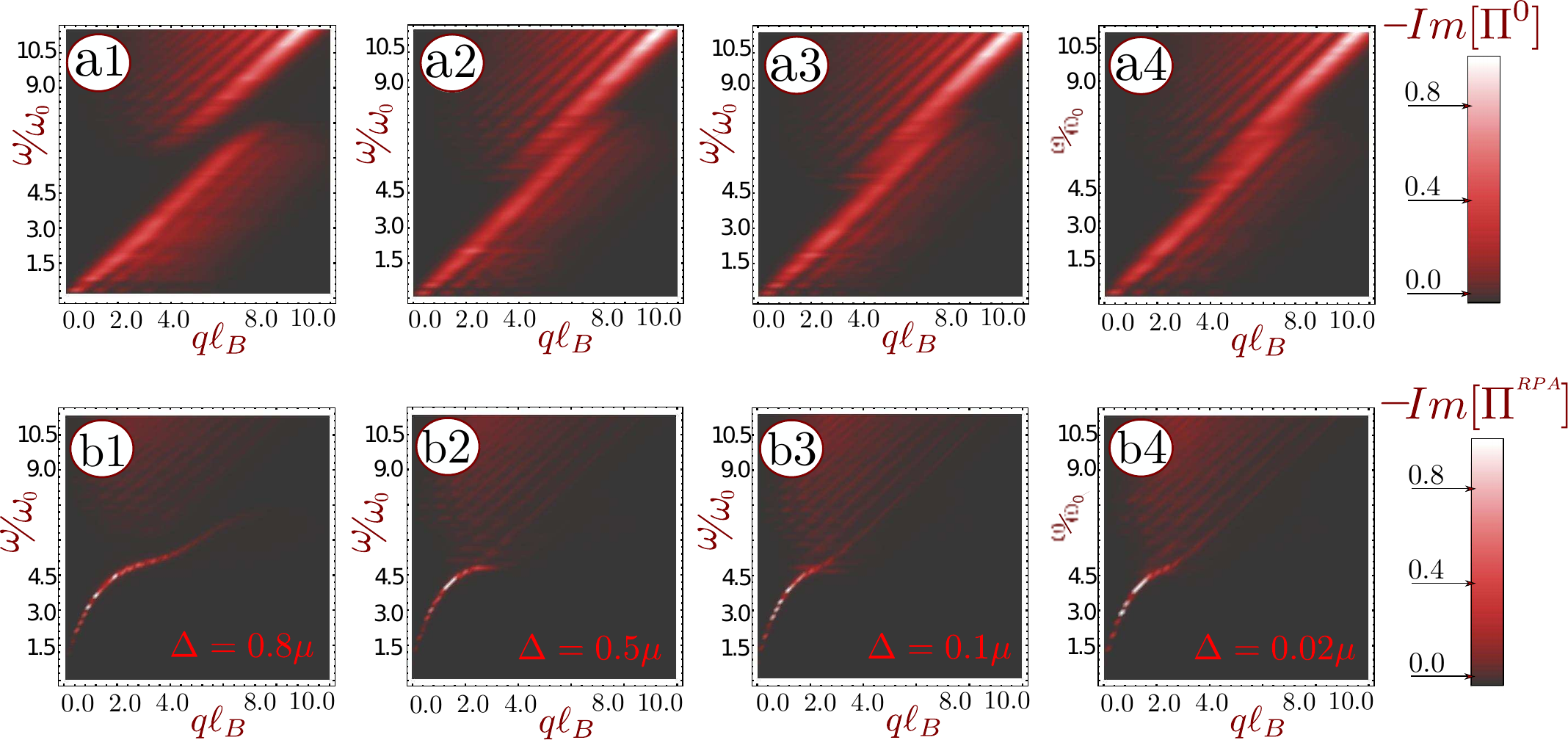}
\caption{Density plots of ${\rm Im}[\Pi^{0}(q,\,\omega)]$ and ${\rm Im}[\Pi^{\rm RPA}(q,\,\omega)]$ for \textbf{gapped graphene}.
The upper panel a(1)-a(4) presents results for the non-interacting polarization function, while the lower panel b(1)-b(4) presents the RPA renormalized ones.
Each plot  in 1 through 4 corresponds to $\Delta/\mu=0.8$, $0.5$, $0.1$ and $0.02$.}
\label{AI-fig-8}
\end{figure}

The particle-hole modes for gapless graphene with Dirac cone dispersion
relation  are presented in Fig.\,\ref{AI-fig-7}. The excitation regions are basically determined by the inclined straight lines obeying $\omega=v_F q$.
A dark triangular region, located near  the origin, indicates the  undamped plasmons. Although non-dispersive particle-hole
modes are not clearly seen, they still alter the particle-hole excitation spectrum at $B=0$.
The damping-free plasmon curve occurring in ${\rm Im}[\Pi^{\rm RPA}(q,\,\omega)]$ follows the $\sqrt{q}$-dispersion in the long-wavelength limit,
switching to a straight line to become damped plasmons. Similar features are found from Figs.\,\ref{AI-fig-7}(b1) and (b2) for higher doping. However, the
increased $\mu$ enlarges the region for undamped plasmons. Moreover, the
RPA renormalization displayed in Figs.\,\ref{AI-fig-7}((b2)-(b4) is most
significant for a larger value of interaction parameter $r_s$.
From Figs.\,\ref{AI-fig-7} and \ref{AI-fig-6}, we also compare the RPA effects (${\rm Im}[\Pi^{\rm RPA}(q,\,\omega)]$) with various values of the interaction
parameter  $r_s$ for graphene and 2DEG. The Coulomb interaction in 2DEG leads
to the existence of dispersive magnetoexcitons. In the limit of vanishing interaction $r_s\to 0$, however, the renormalized
RPA polarization becomes similar to that of non-interacting electrons, as seen from the comparisons of (b1) and
(b4) in Fig.\,\ref{AI-fig-6} as well as in Fig.\,\ref{AI-fig-7}.
Finally, Fig.\,\ref{AI-fig-8} shows the  effect  on the plasmons in graphene, where the undamped plasmon curve does not follow the $v_Fq$ line anymore but
follows  that for $B=0$ instead, leading to a suppression of magnetic field effects.
Collective excitations in AA- and AB-stacked graphene in the presence of magnetic field have been investigated recently by Wu.et.al \cite{bib:AI:Stack}. The groupings of the Landau levels lead to considerable differences in the
plasmon excitation energies, which are determined by the way in which the layers are stacked.
In general, plasmonics in carbon-based nanostructures has recently received considerable development.
First, plasmons modes were caclulated and studied in AA-stacked bilayer graphene \cite{bib:AI:RafaelBi}.
The above mentioned Klein tunneling in this type of bilayer graphene was adressed in \cite{bib:AI:AA}
Plasmon excitations of a single $C_{60}$ molecule, induced by an external, fast moving electron,
were theoretically studied based on quantum hydrodynamical model \cite{bib:AI:Zoran1}. This study
resulted in finding the differential cross sections. The localization of charged particles by the
image potential of spherical shells, such as fullerene buckyballs, has been addressed in \cite{bib:AI:ourI}
Systems without spherical symmetry demonstrate anisotropy and dimerization of the plasmon excitations
\cite{bib:AI:ourp1,bib:AI:ourp2}. Plasmon frequencies of such systems depend not only on the angular 
momentum $L$, but its projection $M$ on the axis of quantization. Collective excitations in the 
processes of photoionization and electron inelastic scattering were addressed in \cite{bib:AI:V}.     

\section{Concluding Remarks and Research Outlook}

In summary, we  reviewed the effect of an energy gap on the electronic,
transport and many-body properties of graphene. The gap may be generated  
by a number of means in graphene and topological insulators. This includes a substrate
or finite width.   However, we paid attention  to the gap  which appears
as a result of electron-photon interaction between Dirac electrons in
graphene and circularly-polarized photons. This type of energy gap  is
tunable and may be varied in experiment with the intensity of the imposed
light.

We note that the presence of the gap in the energy dispersion leads to the
breaking of chiral symmetry. Consequently, this leads to significant changes
in the electron transmission. There is a decrease of the electron transmission
amplitude for a head-on collision and almost head-on incidence  as a result
of the fact that the Klein paradox no longer exists. The transmission resonances
(the peaks for finite angles of incidence) are slightly shifted  although the 
general structure of the peaks remains unaffected. However, if two or more
pairs of subbands are taken into consideration, the transmission may even i
ncrease compared to the case of only one photon.

We showed a minimum mobility before a field threshold for entering into the nonlinear-transport regime due to build-up of a frictional force.
We demonstrated a mobility enhancement after this threshold value because of heated electrons in high energy states.
We also obtained a maximum mobility enhancement due to balance between simultaneously increasing group velocity and phonon scattering.
Additionally, we proved a increased field threshold by a small correlation length for the line-edge roughness.

Also, we considered magnetoplasmons in gapped graphene and concluded that
similar to the case of zero magnetic field, the presence of the gap
increases the region free from particle-hole excitations. Consequently,
the region  where undamped plasmons exist is expanded. We obtained a new
type of plasmon dispersion.

\vspace{0.6in}
\noindent
{\Large{\bf Acknowledgements}}
\vspace{0.2in}
This research was supported by a grant from  the AFOSR and a  contract from the AFRL. We are grateful to Mary Michelle Easter for a critical reading of the manuscript and helpful  comments.


\end{document}